\documentstyle [12pt]{article}
\makeatletter \oddsidemargin  0in \evensidemargin 0in
\textwidth16cm \RequirePackage[dvips]{graphicx} \textheight 20.5cm
\newcommand{\singlespacing}{\let\CS=\@currsize\renewcommand{\baselinestretch}{1.5}\tiny\CS}
\makeatletter \oddsidemargin  -.1in \evensidemargin -.1in
\newcommand{\doublespacing}{\let\CS=\@currsize\renewcommand{\baselinestretch}{1.35}\tiny\CS}

\def\@citex[#1]#2{\if@filesw\immediate\write\@auxout{\string\citation{#2}}\fi
  \def\@citea{}\@cite{\@for\@citeb:=#2\do
    {\@citea\def\@citea{,\linebreak[0]\hskip0pt plus .2em}%
      \@ifundefined{b@\@citeb}%
    {{\bf ?}\@warning{Citation `\@citeb' on page \thepage\space undefined}}%
      \hbox{\csname b@\@citeb\endcsname}}}{#1}}

\newtheorem{rule-def}[theorem]{Rule}
\begin{document}
\title{Inseparability of Quantum Parameters}
\author{I.Chakrabarty $^1$\thanks {indranilc@indiainfo.com},S.Adhikari $^2$ ,Prashant $^3$,B.S.Choudhury $^2$\\
 $^1$ Heritage Institute of Technology, Kolkata, India\\
 $^2$ Bengal Engineering and Science University, Howrah, India\\
$^3$ Indian Institute of IT and Management, India }
\date{}
\maketitle{}
\begin{abstract}
In this work, we show that 'splitting of quantum information' [6]
is an impossible task from three different but consistent
principles of unitarity of Quantum Mechanics, no-signalling
condition and non increase of entanglement under Local Operation
and Classical Communication.
\end{abstract}
PACS Numbers: 03.67.-a, 03.65.Bz, 89.70.+c
\section{\bf Introduction}
In quantum information theory it is most important of knowing the
various differences between the classical and quantum information.
Many operations which are feasible in digitized information
becomes an impossibility in quantum world [1-6]. This may be
probably due to the linear structure or may be due to the unitary
evolution in quantum mechanics. Regardless of their origin, these
impossible operations are making quantum information processing
much more restricted than it's classical counterpart. On the other
hand this restriction on many quantum information processing tasks
is making quantum information more secure. In the famous land mark
paper of Wootters and Zurek it was shown that a single quantum
cannot be cloned [1]. Later it was also shown by Pati and
Braunstein that we cannot delete either of the two quantum states
when we are provided with two identical quantum states at our
input port [2]. In spite of these two famous 'no-cloning' [1] and
'no-deletion' [2] theorem there are many other 'no-go' theorems
like 'no-self replication' [3] , 'no-partial erasure' [5],
'no-splitting' [6] and many more which have come up. Recent
research has revealed that these theorems are consistent with
different principles like principle of no-signalling and
conservation of entanglement under LOCC [7-9]. If we put it in a
different way it means that if we violate these 'no-go' theorems
we will violate the principle of no-signalling and non increase
of entanglement under LOCC. \\
{\bf No-splitting theorem:} It is a well known fact that there are
many operations which are feasible in the classical world but
doesn't hold good in the quantum domain. These we generally refer
as 'General impossible operations'. "No splitting Theorem" [6]
(almost equivalent to the 'No-Partial Erasure of Quantum
Information' [5])is yet another addition to this set. It states
that 'For an unknown qubit,quantum information cannot be split
into two complementing qubits,i.e. the information in one qubit is
an inseparable entity. An important application of splitting of
quantum information is a construction of a gate which can be used
to reversibly  split a parameter encoded in non orthogonal quantum
states , enabling the necessary quantum information compression
and decompression required for optimal quantum cloning with
multiple copies [10].\\
 In this work our objective is different from
[6] in the sense that here we will investigate whether we can
split two non orthogonal quantum states from three different but
consistent principles like, preservation of inner products under
unitary evolution, the principle of non increase of entanglement
under LOCC and principle of no signalling . In other words, unlike
in ref [6], instead proving the splitting of quantum state from
linearity, we correlate this impossibility with other aspects
like, unitarity ,restrictions on entanglement processing and
causality. \\
\section{\bf Proof of No-splitting theorem from three different principles:}
{\bf I. From Unitarity of Quantum Mechanics: } First of all we
show that the 'No-splitting theorem'is consistent with the unitary
evolution of quantum theory. For this purpose we will consider a
pair of non orthogonal states
$[|\psi_1(\theta_1,\phi_1)\rangle,|\psi_1(\theta_2,\phi_2)\rangle]$,
where $0\leq \theta \leq \pi$ and $0\leq \phi \leq 2\pi$. These
non orthogonal states are represented by points on the Bloch
sphere. Let us assume that the splitting of quantum information
into complementary parts is possible. If we consider a
hypothetical machine which can split quantum information in each
of the non-orthogonal states into complementary parts, then the
action of the machine on the non-orthogonal states
$[|\psi_1(\theta_1,\phi_1)\rangle,|\psi_1(\theta_2,\phi_2)\rangle]$
is defined by the set of transformations:
\begin{eqnarray}
|\psi_1(\theta_1,\phi_1)\rangle|\psi_2\rangle\rightarrow|\psi_1(\theta_1)\rangle|\psi_2(\phi_1)\rangle\\
|\psi_1(\theta_2,\phi_2)\rangle|\psi_2\rangle\rightarrow|\psi_1(\theta_2)\rangle|\psi_2(\phi_2)\rangle
\end{eqnarray}
where
\begin{eqnarray}
|\psi_1(\theta_j,\phi_j)\rangle=\cos(\frac{\theta_j}{2})|0\rangle+\sin(\frac{\theta_j}{2})e^{(i\phi_j)}|1\rangle\\
|\psi_2(\phi_j)\rangle= |\psi_{21}\rangle+e^{(i\phi_j)}|\psi_{22}\rangle\\
|\psi_1(\theta_j)\rangle
=\cos(\frac{\theta_j}{2})|\psi_{11}\rangle+\sin(\frac{\theta_j}{2})|\psi_{12}\rangle
\end{eqnarray}
where $j=(1,2)$. Here
$|\psi_{11}\rangle,|\psi_{12}\rangle,|\psi_{21}\rangle,|\psi_{22}\rangle$
are non normalized states independent of $\theta,\phi$. The
unitarity of transformations will preserve the inner product .
\begin{eqnarray}
\langle
\psi_1(\theta_1,\phi_1)|\psi_1(\theta_2,\phi_2)\rangle=\nonumber\\\langle
\psi_1(\theta_1)|\psi_1(\theta_2)\rangle\langle
\psi_2(\phi_1)|\psi_2(\phi_2)\rangle
\end{eqnarray}
The above equality will not hold for all values of
$(\theta,\phi)$. The equality will hold if $\phi_2=\phi_1+n\pi$
and $\theta_1\pm\theta_2=(2m+1)\pi$ where $m,n$ are integers(see
appendix).This equality corresponds to a situation where the
quantum states are orthogonal. Thus we see that the equality does
not hold for all values of $\theta$ and $\phi$ and hence we
conclude that this kind of transformation doesn't exist. We cannot
split the quantum information for two non orthogonal quantum
states.\\\\

 {\bf II. From Non Increase of Entanglement
under LOCC and No signalling Condition:} Next we show that
splitting of quantum information is an impossible operation from
the principle of non increase of entanglement under LOCC. Let us
consider an entangled state shared by two distant parties Alice
and Bob, of the form
\begin{eqnarray}
|\psi\rangle_{AB}=\frac{1}{\sqrt{2}}[|0\rangle_{A}|\psi_1(\theta_1,\phi_1)\rangle_{B}+|1\rangle_{A}|\psi_1(\theta_2,\phi_2)\rangle_{B}]|\psi_2\rangle_{B}
\end{eqnarray}
where $\{|\psi_2\rangle\}$ is the blank state attached to the
Bob's particle. \\
The reduced density matrix on Alice's side is given by,
\begin{eqnarray}
\rho_{A}&=&Tr_{B}(|\psi\rangle_{ABAB}\langle\psi|)=\nonumber\\&&\frac{1}{2}[I+{}|1\rangle\langle0|(\langle
\psi_1(\theta_1,\phi_1)|\psi_1(\theta_2,\phi_2)\rangle)+\nonumber\\&&|0\rangle\langle1|(\langle
\psi_1(\theta_2,\phi_2)|\psi_1(\theta_1,\phi_1)\rangle)]
\end{eqnarray}
 Let us assume that Bob is in possession of a machine which will split the quantum information
 of his particle. The transformation describing the action of the machine is given by equations (1) and
 (2). Now after the application of the quantum information
 splitting machine the entangled state (7) takes the form
\begin{eqnarray}
|\psi\rangle_{AB}^C=\frac{1}{\sqrt{2}}[|0\rangle_{A}|\psi_1(\theta_1)\rangle_{B}|\psi_2(\phi_1)\rangle_{B}+|1\rangle_{A}|\psi_1(\theta_2)\rangle_{B}|\psi_2(\phi_2)\rangle_{B}]
\end{eqnarray}
The reduced density matrix on the Alice's side after the
application of the machine is given by,
\begin{eqnarray}
\rho_{A}^{C}&=&\frac{1}{2}[I+|1\rangle\langle0|(\langle
\psi_1(\theta_1)|\psi_1(\theta_2)\rangle)(\langle
\psi_2(\phi_1)|\psi_2(\phi_2)\rangle){}\nonumber\\&&+|0\rangle\langle1|(\langle
\psi_1(\theta_2)|\psi_1(\theta_1)\rangle)(\langle
\psi_2(\phi_2)|\psi_2(\phi_1)\rangle)]
\end{eqnarray}
The respective largest eigen values of these two reduced density
matrices are given by
\begin{eqnarray}
\lambda_A=\frac{1}{2}+\frac{|p|^2}{2}\\
\lambda_A^C=\frac{1}{2}+\frac{|q|^2|r|^2}{2}
\end{eqnarray}
where $p=\langle
\psi_1(\theta_2,\phi_2)|\psi_1(\theta_1,\phi_1)\rangle$,
$q=\langle \psi_1(\theta_2)|\psi_1(\theta_1)\rangle$, $r=\langle
\psi_2(\phi_2)|\psi_2(\phi_1)\rangle$. To show that the amount of
entanglement $E(|\psi\rangle_{AB})$ and $E(|\psi\rangle_{AB}^C)$
of the respective entangled states before and after the splitting
doesn't increase, we must show that $\lambda_A<\lambda_A^C$. To
show $\lambda_A<\lambda_A^C$ this  we must show that,
\begin{eqnarray}|\langle \psi_1(\theta_2,\phi_2)|\psi_1(\theta_1,\phi_1)\rangle|<\nonumber\\|\langle
\psi_1(\theta_2)|\psi_1(\theta_1)\rangle||\langle
\psi_2(\phi_2)|\psi_2(\phi_1)\rangle|
\end{eqnarray}
\begin{eqnarray}
&&LHS:|\cos(\frac{\theta_1}{2})\cos(\frac{\theta_2}{2})+e^{i(\phi_1-\phi_2)}\sin(\frac{\theta_1}{2})\sin(\frac{\theta_2}{2})|
\nonumber\\{}
 &&RHS:|\cos(\frac{\theta_1}{2})\cos(\frac{\theta_2}{2})+\sin(\frac{\theta_1}{2})\sin(\frac{\theta_2}{2})|\nonumber\\{}&&|1+e^{i(\phi_1-\phi_2)}|\end{eqnarray}
Let
\begin{eqnarray}
(\phi_1-\phi_2)= k,
\cos(\frac{\theta_1}{2})\cos(\frac{\theta_2}{2})=x \\
\sin(\frac{\theta_1}{2})\sin(\frac{\theta_2}{2})=y
\end{eqnarray}
where x and y are real quantities. \\
Now $|\langle
\psi_1(\theta_2,\phi_2)|\psi_1(\theta_1,\phi_1)\rangle|=|x+e^{ik}y|=|[x+y\cos(k)]+iy[\sin(k)]|=\sqrt{[x^2+y^2+2xy\cos(k)]}$
and $|\langle \psi_1(\theta_2)|\psi_1(\theta_1)\rangle||\langle
\psi_2(\phi_2)|\psi_2(\phi_1)\rangle|=(x+y)\sqrt{2(1+\cos(k))}$.\\

Therefore $A$(say)=$[|\langle
\psi_1(\theta_2,\phi_2)|\psi_1(\theta_1,\phi_1)\rangle|]^2
-[|\langle \psi_1(\theta_2)|\psi_1(\theta_1)\rangle||\langle
\psi_2(\phi_2)|\psi_2(\phi_1)\rangle| ]^2
=x^2+y^2+2xy\cos(k)-2(x+y)^2(1+\cos(k))=-[x^2+y^2+4xy+2(x^2+y^2+xy)\cos(k)]
>0$
for some values of x,y,k. This implies that, $\lambda_A>
\lambda_A^C\Rightarrow
E(|\psi\rangle_{AB}^C)>E(|\psi\rangle_{AB})$, as a consequence of
which we can say that the amount of entanglement will increase
under local operation. However we know that entanglement is non
increasing under such operations [ in general one can only claim
that it is conserved under a bilocal unitary operation ]. This
gives rise to contradiction. Therefore, it is clear that the
principle of non increase of entanglement under LOCC doesn't allow
perfect splitting of nonorthogonal quantum states. This rules out
the existence of a hypothetical quantum information splitting
machine, designed to split the quantum information of a
nonorthogonal quantum state
.\\\\

Next we show that the splitting of quantum information is not
possible from the principle of no signalling. In other words we
can say that if we assume perfect splitting of quantum information
it will violate the principle of no-signalling.\\
Suppose we have a singlet state shared by two distant parties
Alice and Bob . The singlet state can be written in two different
basis as
\begin{eqnarray}
|\chi\rangle=\frac{1}{\sqrt{2}}(|\psi_1\rangle|\overline{\psi_1}\rangle-|\overline{\psi_1}\rangle|\psi_1\rangle)\nonumber\\
=\frac{1}{\sqrt{2}}(|\psi_2\rangle|\overline{\psi_2}\rangle-|\overline{\psi_2}\rangle|\psi_2\rangle)
\end{eqnarray}
where $\{|\psi_1\rangle, |\overline{\psi_1}\rangle \}$ and $\{
|\psi_2\rangle, |\overline{\psi_2}\rangle \}$ are two sets of
mutually orthogonal spin states (qubit basis). Alice possesses the
first particle while Bob possesses the second particle. Alice can
choose to measure the spin in any one of the qubit basis namely
$\{|\psi_1\rangle, |\overline{\psi_1}\rangle \}$, $\{
|\psi_2\rangle, |\overline{\psi_2}\rangle \}$. The theorem of no
signalling tells us that the measurement outcome of Bob are
invariant under local unitary transformation done by Alice on her
qubit.The density matrix $ \rho_{B}= tr\rho_{AB}= tr[(U_{A}\otimes
I_{B})\rho_{AB}(U_{A}\otimes I_{B})^{\dagger}]$ is invariant under
local unitary operation by Alice . Hence Bob cannot distinguish
two mixtures due to the unitary operation done at remote place.
One may ask if Bob split the quantum information of his particle
and if Alice measure her particle in either of the two basis then
is there any possibility that Bob know the basis in which Alice
measure her qubit or in other words, is there any way by which Bob
using a perfect splitting machine can distinguish the statistical
mixture in his subsystem resulting from the measurement done by
Alice. If Bob can do this then signalling will take place, which
is impossible. Hence now our task is to show that the splitting of
information
is an impossible task from no-signalling principle.  \\
Let us consider a situation where Bob is in possession of a
hypothetical quantum information splitting machine. The unitary
transformation describing the splitting of quantum information for
an input state $|\psi_i(\theta,\phi)\rangle$ (where i=1,2) is
defined as ,
\begin{eqnarray}
|\psi_i(\theta,\phi)\rangle|\Sigma\rangle\rightarrow|\psi_i(\theta)\rangle|\Sigma(\phi)\rangle\nonumber\\
|\overline{\psi_i(\theta,\phi)}\rangle|\Sigma\rangle\rightarrow|\overline{\psi_i(\theta)}\rangle|\overline{\Sigma(\phi)}\rangle
\end{eqnarray}
where $\{|\Sigma\rangle\}$ is the ancilla state attached by Bob .\\
After the application of the transformation defined in (18) by Bob
on his particle the singlet state defined by (17) including the
ancilla state attached by Bob reduces to the form,
\begin{eqnarray}
&&|\chi\rangle|\Sigma\rangle\rightarrow
|\chi\rangle^{S}={}\nonumber\\&&
\frac{1}{\sqrt{2}}[|\psi_1(\theta,\phi)\rangle|\overline{\psi_1(\theta)}\rangle|\overline{\Sigma(\phi)}\rangle
{}\nonumber\\&&-|\overline{\psi_1(\theta,\phi)}\rangle|\psi_1(\theta)\rangle|\Sigma(\phi)\rangle]={}\nonumber\\&&
\frac{1}{\sqrt{2}}[|\psi_2(\theta,\phi)\rangle|\overline{\psi_2(\theta)}\rangle|\overline{\Sigma(\phi)}\rangle
{}\nonumber\\&&-|\overline{\psi_2(\theta,\phi)}\rangle|\psi_2(\theta)\rangle|\Sigma(\phi)\rangle]
\end{eqnarray}
After Bob applying the splitting machine on his qubit, Alice can
measure her particle in two different basis. If Alice measures her
particle in the basis $\{|\psi_1\rangle, |\overline{\psi_1}\rangle
\}$, then the reduced density matrix in the Bob's subsystem
(including ancilla) is given by,
\begin{eqnarray}
&&\rho_{BC}=tr_{A}(\rho_{ABC}){}\nonumber\\&&
=\frac{1}{2}\{|\overline{\psi_1(\theta)}
\overline{\Sigma(\phi)}\rangle\langle\overline{\psi_1(\theta)}
\overline{\Sigma(\phi)}|{}\nonumber\\&&+|\psi_1(\theta)\Sigma(\phi)\rangle\langle\psi_1(\theta)\Sigma(\phi)|\}
\end{eqnarray}
On the other hand if Alice measures her particle in the basis
$\{|\psi_2\rangle, |\overline{\psi_2}\rangle \}$ then the state
described by the reduced density matrix in the Bob's side is given
by,
\begin{eqnarray}
&&\rho_{BC}=tr_{A}(\rho_{ABC}){}\nonumber\\&&
=\frac{1}{2}\{|\overline{\psi_2(\theta)}
\overline{\Sigma(\phi)}\rangle\langle\overline{\psi_2(\theta)}
\overline{\Sigma(\phi)}|{}\nonumber\\&&+|\psi_2(\theta)\Sigma(\phi)\rangle\langle\psi_2(\theta)\Sigma(\phi)|\}
\end{eqnarray}
Since the statistical mixture in (20) and (21) are different, so
this would have allowed Bob to distinguish in which basis Alice
has performed measurement, thus allowing for super luminal
signalling. However the criterion of 'No-signalling' tells us that
communication faster than light is not possible. So we arrive at a
contradiction, that is, the transformation defined in (18) is not
possible in quantum world. This rules out the existence of
hypothetical machine like quantum information splitting machine.\\
\section{ \bf Conclusion} In this work we show that the total information
contained in the quantum state on the Bloch-sphere cannot be
written as the tensor product of the state containing the
information of the azimuthal angle and the state containing the
information of the phase angle. Therefore the quantum information
can be regarded as an inseparable entity or in other words we can
say that it is impossible to express the function of $\theta$ and
$\phi$ as the product of function of $\theta$ alone and function
of $\phi$ alone. To justify the above statement, we proved the
no-splitting theorem from three different principles: (i)
Unitarity of quantum mechanics (ii) principles of non increase of
entanglement under LOCC and (iii) Principle of no signalling.

\section{\bf Acknowledgement}
I.C acknowledges Prof C.G.Chakraborti, S.N.Bose Professor of
Theoretical Physics, Department of Applied Mathematics, University
of Calcutta for being the source of inspiration in carrying out
research. S.A acknowledges CSIR project no.F.No.8/3(38)/2003-EMR-1
for providing financial support . I.C and S.A would like to thank
Dr.A.K.Pati for useful discussion.
\section{\bf Appendix}
$\langle
\psi_1(\theta_1,\phi_1)|\psi_1(\theta_2,\phi_2)\rangle=\cos\frac{\theta_1}{2}\cos\frac{\theta_2}{2}
+e^{i(\phi_2-\phi_1)}\sin\frac{\theta_1}{2}\sin\frac{\theta_2}{2}$
, $\langle
\psi_1(\theta_1)|\psi_1(\theta_2)\rangle=\cos\frac{\theta_1}{2}\cos\frac{\theta_2}{2}+\sin\frac{\theta_1}{2}\sin\frac{\theta_2}{2}$,
and $\langle
\psi_2(\phi_1)|\psi_2(\phi_2)\rangle=1+e^{i(\phi_2-\phi_1)}$. Now
we have to show that, $\langle
\psi_1(\theta_1,\phi_1)|\psi_1(\theta_2,\phi_2)\rangle=\langle
\psi_1(\theta_1)|\psi_1(\theta_2)\rangle\langle
\psi_2(\phi_1)|\psi_2(\phi_2)\rangle$. This equality is possible
only when
\begin{eqnarray}
\tan\frac{\theta_1}{2}\tan\frac{\theta_2}{2}=-e^{i(\phi_2-\phi_1)}
\end{eqnarray}
Now by equating the real and imaginary parts of the above
expression we get,
\begin{eqnarray}
\tan\frac{\theta_1}{2}\tan\frac{\theta_2}{2}=\cos(\phi_2-\phi_1)
\end{eqnarray}
\begin{eqnarray}
\sin(\phi_2-\phi_1)=0
\end{eqnarray}
On simplifying equation (19) we get, $(\phi_2-\phi_1)=n\pi$ ,where
$n=0,\pm1,\pm2,.....$. Using $(\phi_2-\phi_1)=n\pi$ equation (17)
reduces to the form
$\tan\frac{\theta_1}{2}\tan\frac{\theta_2}{2}=(-1)^{n+1}$.\\
Now we consider two cases:\\
\textbf{Case1.} When n is even,
$\tan\frac{\theta_1}{2}\tan\frac{\theta_2}{2}=-1\Rightarrow
(\theta_1-\theta_2)=(2m+1)\pi$.\\ \textbf{Case2.}
 When n is odd,
$\tan\frac{\theta_1}{2}\tan\frac{\theta_2}{2}=1\Rightarrow
(\theta_1+\theta_2)=(2m+1)\pi$. (where, $m=0,\pm1,\pm2,.....$)
Thus we see that the unitarity of the transformation is preserved
only when $\phi_2=\phi_1+n\pi$ and
$\theta_1\pm\theta_2=(2m+1)\pi$.
\section{\bf Reference}
$[1]$ W.K.Wootters and W.H.Zurek,Nature \textbf{299} (1982) 802-803.\\
$[2]$ A.K.Pati and S.L.Braunstein Nature \textbf{404} (2000) 164.\\
$[3]$ A.K.Pati and S.L.Braunstein, Quantum mechanical universal
constructor,quantph/0303124 (2003).\\
$[4]$ A.K.Pati, Phys. Rev. A \textbf{66}, 062319(2002).\\
$[5]$ A.K.Pati and Barry C. Sanders, No partial erasure of quantum
information,quant-ph/0503138\\
$[6]$ D.Zhou.et.al, Quantum information cannot be split into
complementary parts,quant-ph/0503168\\
$[7]$ A.K.Pati and S.L.Braunstein,Phys.Lett.A \textbf{315},208-212 (2003)\\
$[8]$ I.Chattopadhyay.et.al.Phys. Lett. A, \textbf{351},
384-387 (2006).\\
$[9]$ I.Chakrabarty,A.K.Pati,S.Adhikari,B.S.Choudhury, General
Impossible Operations as a consequence of principles of quantum
information (under preparation).\\
$[10]$ A.Chefles and S.M. Barnett , Phys. Rev. A \textbf{60},136
(1999).
\end{document}